\documentclass[a4paper]{jpconf}
\usepackage{graphicx}
\usepackage{iopams}
\usepackage [latin1]{inputenc}
\usepackage{color}
\usepackage{citesort}
\newcommand{\comment}[1]{}

\begin{document}

\title{Multiband character of $\beta$-FeSe: Angular dependence of the magnetoresistance and upper critical field }

\author{M. L. Amig\'o$^1$, V. Ale Crivillero$^1$, D. G. Franco$^{1}$, and G. Nieva$^{1,2}$}

\address{$^1$ Comisi\'on Nacional de Energ\'{i}a At\'omica - Centro At\'omico
Bariloche and Instituto Balseiro, Universidad Nacional de Cuyo, S.C. de Bariloche, Argentina

$^2$ CICT - Consejo Nacional de Investigaciones Cient\'{i}ficas y
T\'ecnicas, Argentina}

\ead{gnieva@cab.cnea.gov.ar}

\begin{abstract}
We studied $ab$-plane transport properties in single crystals of
the superconductor $\beta$-FeSe up to 16~T. In the normal state,
below 90 K,  the crystals present a strongly anisotropic positive
magnetoresistance that becomes negligible above that temperature.
 In the superconducting state (T$_c$=8.87(5) K) the upper critical field anisotropy $H$$_{c2}$$\parallel$$ab$/ $H$$_{c2}$$\parallel$$c$ changes with temperature and the angular dependence of the dissipation for fixed temperatures and fields reflects a strongly anisotropic behavior.
Our results make evident that multiband effects are needed to
describe the measured transport properties. We model the
magnetoresistance and upper critical field behavior with a
two-band model showing that the diffusivities ratio parameter
remains unchanged going from the normal to the superconducting
state.

\end{abstract}

\section{Introduction}

Several parent compounds of the families 1111 and 122 of the Fe based superconductors present a structural transition and spin density wave at similar temperatures. By doping this compounds the transition temperature decreases and superconductivity emerges\cite{RevModPhys.83.1589}.
In the case of pure $\beta$-FeSe there is a structural transition at $T_s\sim$90 K, not accompanied by spin density wave and the samples show superconductivity at low temperatures.  In this compound the orthorhombic distortion and superconductivity do not compete\cite{PhysRevB.87.180505}.
For the Fe-based superconductors, several disconnected Fermi-surface sheets contribute to superconductivity, as revealed by angle-resolved photoemission spectroscopy. In particular for  $\beta$-FeSe these experiments show a significant change in the electronic bands going through the structural transition\cite{2014arXiv1407.1418}. Evidence of the multiband behavior of this material was found in the physical properties such as the magnetic penetration depth\cite{PhysRevLett.104.087004}, the upper critical field, $H_{c2}$, or the Hall coefficient\cite{PhysRevB.85.094515}.
In this work, we grow single crystals of the  $\beta$-FeSe  phase and study the consequences of the structural transition in the normal and superconducting states.

\section{Crystal characterization}

Fe$_{1-y}$Se single crystals were grown using KCl: 2AlCl$_3$ flux in a temperature gradient (hot part of the ampule at 395$^\circ$C and cold part at 385$^\circ$C), for 45 days\cite{C2CE26857D}. This low temperature process allows us to grow the desired phase avoiding high temperature structural transitions and decompositions. We use X Ray diffraction and Energy Dispersive Spectroscopy to characterize the structure and the composition of the single crystals. The crystals have a plate-like shape with the $c$ axis oriented perpendicular to the crystal plane. The crystals have only the tetragonal  $\beta$-FeSe phase present. The lattice parameters are $c=(5.52 \pm 0.01)$\r{A}  and $a=(3.77 \pm 0.01)$\r{A},  in good agreement with the literature\cite{PhysRevB.79.014522}. The composition of the samples present a slight Fe deficiency, $y=0.04$.

\section{Results and discussion}

\begin{figure}[b]
\begin{minipage}{18pc}
\includegraphics[width=18pc]{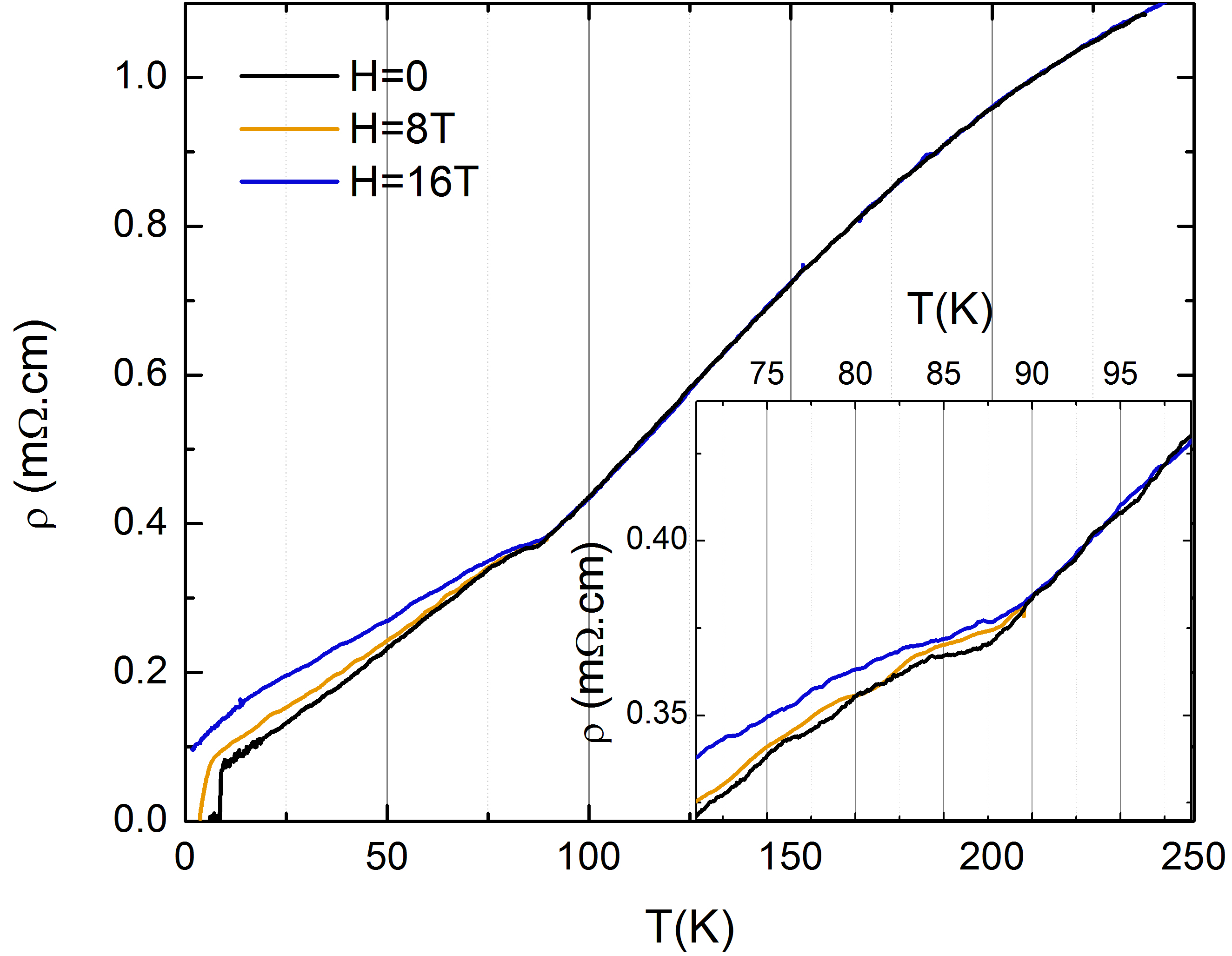}
%\vspace{0.3cm}
\caption{\label{RvsT}Temperature dependence of the $ab$-plane resistivity for a $Fe_{0.96}Se$ crystal  with applied magnetic field parallel to the $c$ axis ($ H=$ 0, 8 and 16 T). $Inset:$ Detail of the resistivity near the structural transition.}
\end{minipage}\hspace{1pc}
\begin{minipage}{18pc}
\begin{center}
\includegraphics[width=18pc]{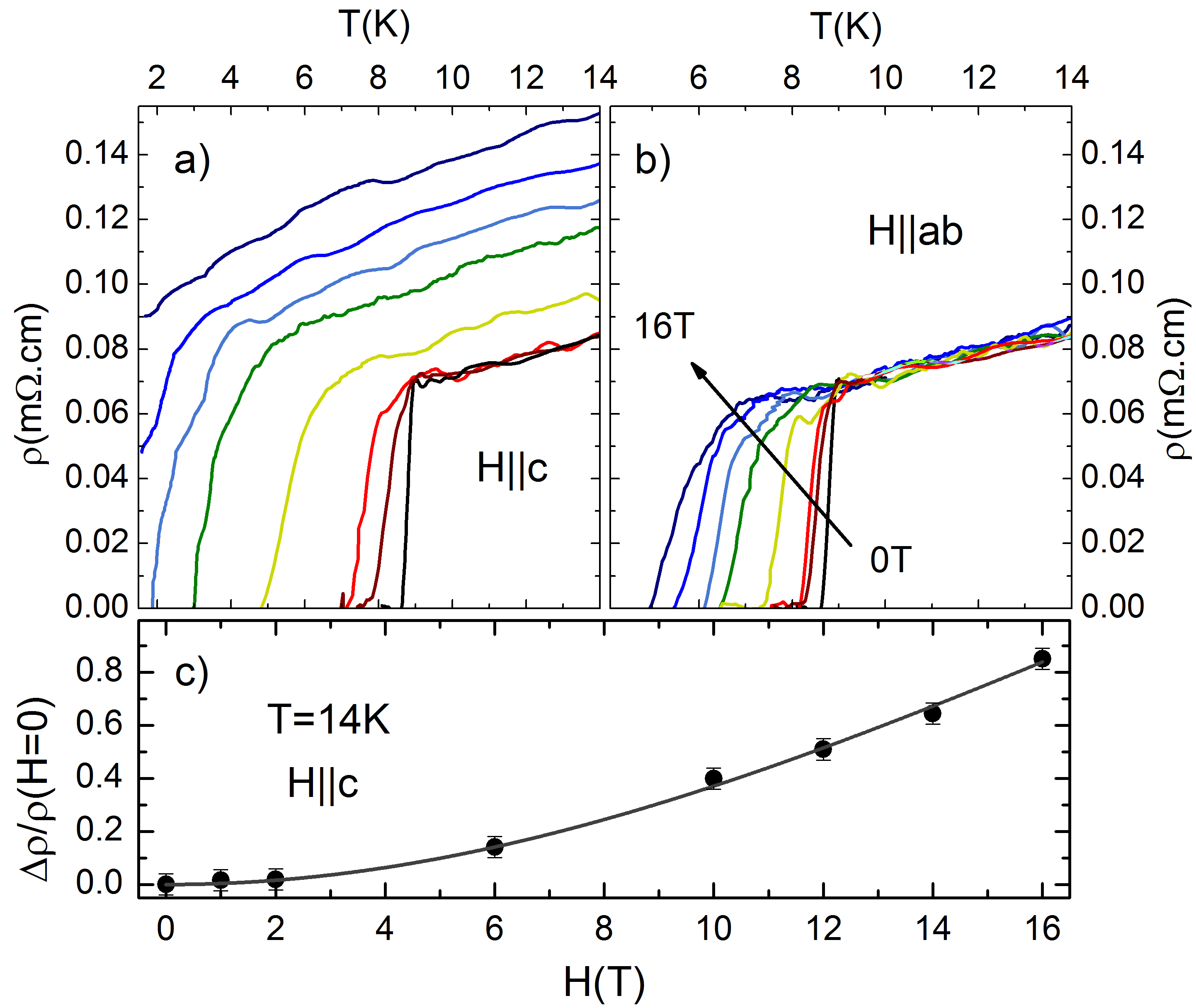}
\caption{\label{RvsT-H}Temperature dependence of the $ab$-plane resistivity for an applied magnetic field ($H=$ 0, 1, 2, 6, 10, 12, 14 and 16 T) $a)$ parallel to the $c$ axis and $b)$ perpendicular to
the $c$ axis. $c)$ Magnetoresistance as a function of the applied magnetic field parallel to the $c$ axis for $T=$ 14 K.}
\end{center}
\end{minipage}
\end{figure}

Figure \ref{RvsT} shows the $ab$-plane resistivity for a single crystal of Fe$_{0.96}$Se with applied magnetic field parallel to the $c$ axis ($H$$\parallel$$c$ = 0, 8 and 16 T). The onset of the transition temperature for $H=0$ is $T_c=(8.87\pm0.05)$~K. The resistivity presents a metallic-like behavior in the normal state and a change in the slope around 90 K. At this temperature there is a structural transition from a tetragonal to an orthorhombic phase\cite{PhysRevLett.103.057002}. Below 90 K we find a positive magnetoresistance when the field is parallel to the $c$ axis, suggesting that its origin is related to the structural transition. A magnetic characterization of the crystals in the normal state shows no evidence of a magnetic feature at this temperature.  In the literature there are similar magnetoresistance results  in films\cite{0953-2048-25-3-035004} and policrystals\cite{magnetorresistencia-policristales}.

In order to study the connection between the electronic properties
of  $\beta$-FeSe above and below the superconducting temperature
we performed electrical transport measurements with an applied
magnetic field up to 16 T parallel and perpendicular to the $c$
axis, see figure \ref{RvsT-H}$a$ and $b$. The electrical contact
configuration on the sample allows the current to flow in the $ab$
plane, always perpendicular to the magnetic field. For $H=0$ there
is a sharp transition, and the transition width slightly increases
with field. Figure  \ref{RvsT-H}$c$ shows the magnetoresistance at
14 K, approximately 5 K above $T_c$ for the field parallel to the
$c$ axis. For field parallel to the $ab$ plane the
magnetoresistance is  negligible as can be seen in figure
\ref{RvsT-H}$b$.
 Moreover, there is a strong difference in the temperature of the superconducting transition onset (i.e. in the upper critical field, $H_{c2}$vs.$T$) for both field directions, indicating an anisotropic material.

Considering that the multiband electronic structure,
characteristic of Fe-based superconductors, was found in many
features of $\beta$-FeSe, we choose a two-band  model to describe
the magnetoresistance data in figure  \ref{RvsT-H}$c$.  Assuming a
spherical symmetry in each band\cite{ziman1965principles}, the
magnetoresistivity for $H$$\parallel$$c$ is given by,
\begin{eqnarray}\label{mr}
\frac{\Delta
\rho}{\rho(H=0)}&=&\frac{\rho(H)-\rho(H=0)}{\rho(H=0)}=\frac{\sigma_1
\sigma_2 (D_1 - D_2)^2H^2}{(\sigma_1+ \sigma_2)^2
+H^2(D_1\sigma_1+D_2\sigma_2)^2}
\nonumber \\
\nonumber \\
 &=&\frac{\sigma_1(\rho(H=0)^{-1}-\sigma_1)D_1^2(1-\eta)^2H^2}{\rho(H=0)^{-2}+H^2D_1^2 \left[ \sigma_1+\eta (\rho(H=0)^{-1}-\sigma_1) \right]^2} ,
\end{eqnarray}
where $\sigma_m$ and $D_m$ are the conductivity and diffusivity in
the $m$ band, $\eta=D_2/D_1$ and
$\rho(H=0)=(\sigma_1+\sigma_2)^{-1}$. For an arbitrary direction
of the magnetic field, $D_m$ must be replaced by
\begin{equation}\label{DvsTh}
D_m(\theta)=\sqrt{(D_m^a)^2cos^2(\theta)+D_m^aD_m^csin^2(\theta)},
\end{equation}
where $\theta$ is the angle between the applied magnetic field and the $c$ axis and $D_m^a$ and $D_m^c$ are the $m$ band diffusivities in plane and along the $c$ axis respectively\cite{PhysRevB.78.174523}.

The magnetoresistance is temperature dependent as can be observed
in figure \ref{RvsT}. For the data  at 14 K,  in figure
\ref{RvsT-H}$c$ ($H$$\parallel$$c$)  the best  fitted curve using
equation \ref{mr}  is shown with a solid line. We used
$\rho$($H=0, T=14$ K)=0.0835 m$\Omega$.cm obtained from figure
\ref{RvsT-H}$a$ and the fit yields:
$\eta(0^\circ)=D_2^a/D_1^a=0.051$, $D_1^a=$0.292 T$^{-1}$ and
$\sigma_1=$36 m$\Omega^{-1}$.cm$^{-1}$. For $H$$\parallel$$ab$
there is a negligible magnetoresistance, which could be explained
if the diffusivities are equal in both bands,
$\eta(90^\circ)=\sqrt{\frac{D_2^aD_2^c}{D_1^aD_1^c}}=1$.

\begin{figure}[h]
\begin{minipage}{18pc}
\includegraphics[width=19pc]{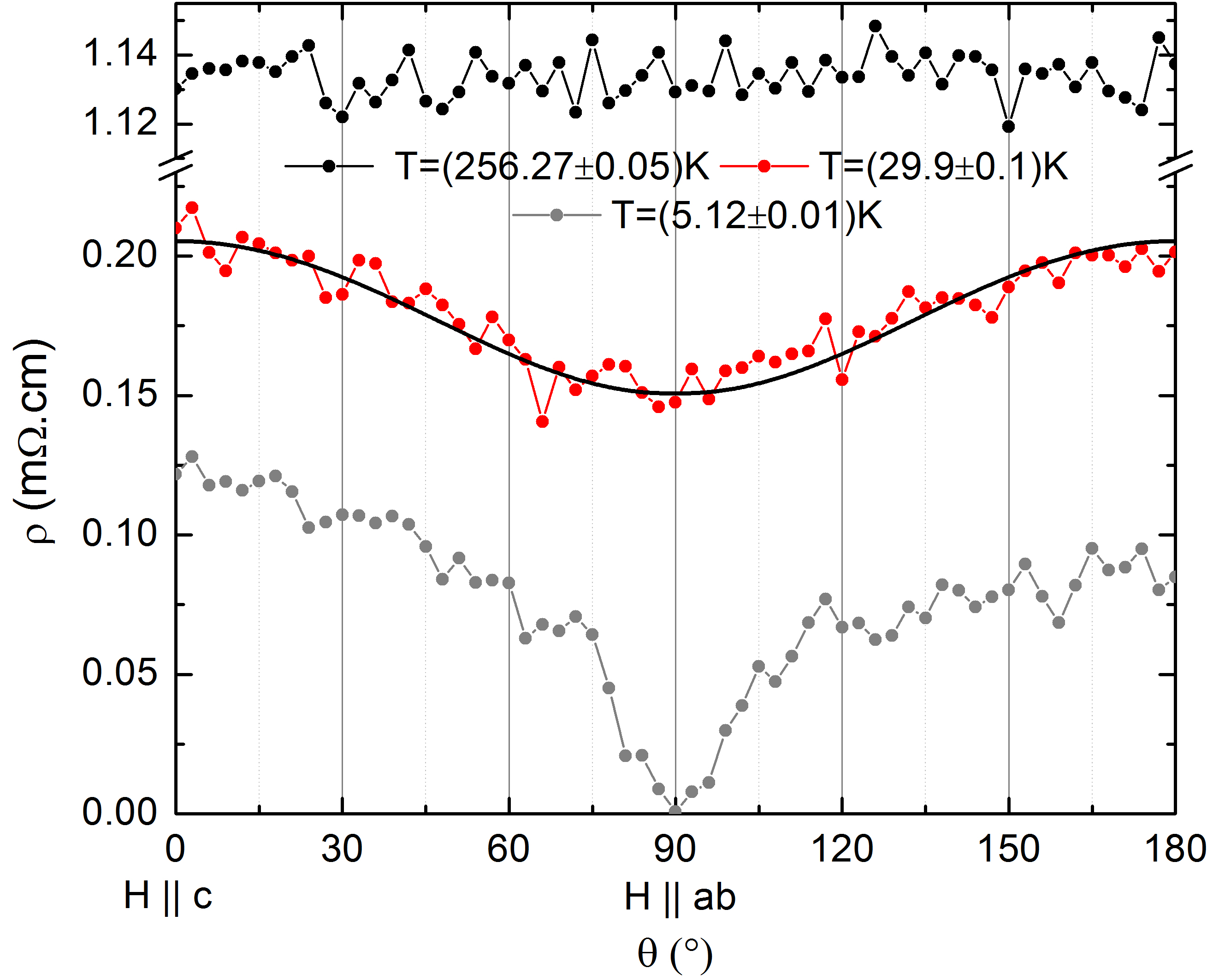}
\vspace {0.8 cm} \caption{\label{RvsTh}$ab$-plane resistivity as a function of the angle between the $c$ axis and the applied magnetic field ($H=$16 T) for different temperatures. The solid line is a fit to the data using equation \ref{mr} and \ref{DvsTh}.}
\end{minipage}\hspace{0.5pc}
\begin{minipage}{18pc}
\begin{center}
\includegraphics[width=17pc]{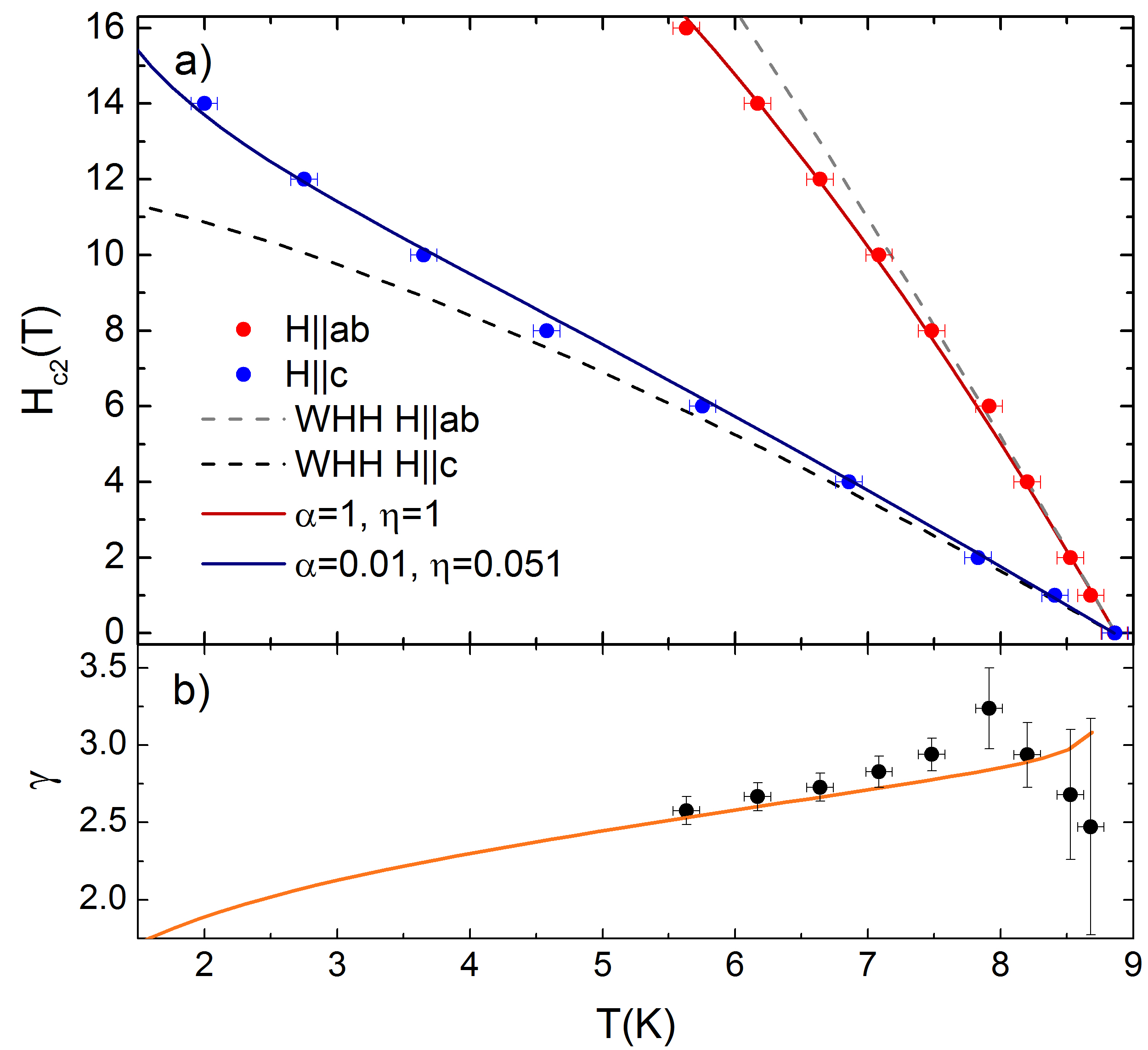}
\caption{\label{Hc2}$a)$ H$_{c2}$(T) for field parallel to the $c$ axis and to the $ab$ plane. In dotted lines is the WHH model ($\eta=1$) with $\alpha=0$ for both field directions. The solid lines are the fit of the equation \ref{U}. $b)$ Anisotropy as a function of temperature and in continuous line is the ratio of the fit for the two directions of the magnetic field.}
\end{center}
\end{minipage}
\end{figure}

We measure the angular dependence of the resistivity at $H=$16 T
for several temperatures. For temperatures above 90 K there is no
angular dependence in accordance with the negligible
magnetoresistance at high temperature. Below the structural
transition temperature there is a smooth angular dependence, see
figure \ref{RvsTh}. Below  $T_c$($H=$0) there is a narrow angular
range in which the sample is superconducting, see the data for
$T=$(5.12 $\pm$ 0.01) K in figure  \ref{RvsTh}. Combining equation
\ref{mr} and \ref{DvsTh} we fitted the data for $T=$(29.9$\pm$0.1)
K. We use
 the same parameters already obtained for $D_1^a$, $D_2^a$ and $\eta(90^\circ)$. The remaining fitted parameters are $\sigma_1$=7.47$m\Omega^{-1}.cm^{-1}$ and $D_1^c=1.62 . 10^{-4}$T$^{-1}$.

According to the model proposed here, there must be a change in the diffusivities at $T\sim$90 K for the emergence of magnetoresistivity, so there must be a change in the effective masses and therefore in the band structure. Recently, some authors studied the changes in the electronic bands at different
temperatures going through the structural transition \cite{2014arXiv1407.1418}\cite{2014arXiv1404.0857N}\cite{PhysRevB.89.220506}.
There is no general agreement about a band reconstruction or variation due to symmetry change at the
 transition. However, such a change in the electronic band structure could be the microscopic
cause of the anisotropic magnetoresistance.

Figure \ref{Hc2}$a$ shows the temperature dependence of the critical field data,  obtained from the onset of the superconducting transition for both field directions. We also plotted the  Werthamer Helfand and Hohenberg (WHH)\cite{PhysRev.147.295} model (dotted line). We observe that this model departs from the experimental data. For $H$$\parallel$$ab$, the data curve towards lower values of $H$, which might indicate the presence of a finite spin paramagnetic effect contribution, that reduces the energy of the normal state, decreasing the value of  $H_{c2}$(T)\cite{PhysRev.147.295}. However, for the case of $H$$\parallel$$c$ the corresponding WHH curve is below the experimental data. This could not be explained introducing a spin paramagnetic effect contribution. A multiple-band scenario must be taken in consideration.
We will describe the superconducting critical field  H$_{c2}$(T) in a two-band scenario. In the dirty limit and including the spin paramagnetic effect, a two-band model yields\cite{PhysRevB.78.174523}
\begin{equation}\label{U}
ln(t)+\frac{1}{2}\Big(U_1(h)+U_2(
h)+\frac{\lambda_0}{w}\Big)-\frac{s}{4}\sqrt{\Big(U_1(h)-U_2(
h)-\frac{\lambda_-}{w^2}\Big)^2+\frac{\lambda_{12}\lambda_{21}}{w^2}}=0,
\end{equation}
\begin{equation}
h=\frac{H_{c2}(T)}{-\frac{dH_{c2}}{dt}\big |_{t=1}},
\end{equation}
\begin{equation}
U_1(h)=Re\bigg(\psi\Big(\frac{1}{2}+(1+i\alpha)\frac{2h}{\pi^2t}\Big)-\psi\Big(\frac{1}{2}\Big)\bigg),
\qquad
U_2(h)=Re\bigg(\psi\Big(\frac{1}{2}+(1+i\frac{\alpha}{\eta})\frac{2h
\eta}{\pi^2t}\Big)-\psi\Big(\frac{1}{2}\Big)\bigg),
\end{equation}
where, $t=T/T_c$, $\lambda_-=\lambda_{11}-\lambda_{22}$, $\lambda_0=\sqrt{\lambda_-^2+4\lambda_{12}\lambda_{21}}$, $w=\lambda_{11}\lambda_{22}-\lambda_{12}\lambda_{21}$, $s=sign(w)$ and $\psi(x)$ is the digamma function. $\lambda_{11}$ and $\lambda_{22}$ are the intraband coupling constant in bands 1 and 2, $\lambda_{12}$ and $\lambda_{21}$ quantify the interband coupling. $\alpha(\theta)=D_0/D_1(\theta)$ takes into account the spin paramagnetic effect, $D_0=\hbar/2m$. Taking $\eta=1$ and $\alpha=0$  we recover the one band WHH model, without the spin orbit coupling.

To further investigate multiband characteristics in $H_{c2}$(T) and the relation with the parameters found in the normal state we fit equation \ref{U} for both field directions (solid lines in figure \ref{Hc2}$a$). We use the $\eta$ values obtained from equation \ref{mr}, $\eta$(0$^\circ$)=0.051 and $\eta$(90$^\circ$)=1. The fitted parameters are $\lambda_{11}=0.13$, $\lambda_{22}=0.1$, $\lambda_{12}=\lambda_{21}=0.0087$, $\alpha$(0$^\circ$)=0.01 and $\alpha$(90$^\circ$)=1. Similar values of $\lambda_{ij}$ are obtained for FeSe in the literature\cite{PhysRevB.85.094515}.

The value of $H_{c2}$(0) strongly depends on $\alpha$, so to give a reliable estimation, it is necessarily to measure at low temperatures and high fields. The values of the parameters obtained here are phenomenological and they are used to show that FeSe presents a multiple band behavior, that is necessary take into account the spin paramagnetic effect and that the different values of the diffusivities could explain the $H_{c2}(T)$ and the magnetoresistivity.
 Figure \ref{Hc2}$b$ present the effective anisotropy ($\gamma= H_{c2}\parallel$$ab / H_{c2}\parallel$$c$). It has a maximum near $T_c$ and decreases with temperature as seen in
others works \cite{PhysRevB.87.134512}\cite{PhysRevB.81.094518}.
This effective anisotropy behavior has  also been taken as an
indication of a multiband character of  $H_{c2}(T)$.

\section{Conclusion}
The  angular dependent magnetoresistance of  $\beta$-FeSe crystals
as well as the anisotropic upper critical field can be described
by a two-band model with parameters such as the band diffusivities
ratio that are valid in a wide temperature range.  In the normal
state, the kink in the $ab$ plane resistivity at 90 K marks the
onset of a very anisotropic magnetoresistance. Above this
temperature, coincident with an structural transition, the
magnetoresistance is negligible.

\ack{We thank Conicet PIP11220090100448 and Sectyp U.N.Cuyo  for financial support.}

\section*{References}
\bibliography{mibib}{}
\bibliographystyle{iopart-num}

\end{document}